\newcommand{\bea}{\begin{eqnarray}}
\newcommand{\eea}{\end{eqnarray}}
\documentclass[twocolumn,prl,preprintnumbers,amsmath,amssymb,superscriptaddress]{revtex4}
\usepackage{graphicx}
\def\6#1{{\underline{#1}}}
\def\m6#1{{\underline{#1}\,}}

\newdimen\Tdim
\def\ispan{{\setbox0=\hbox{i}%
\Tdim\ht0\advance\Tdim\dp0\rule[-\dp0]{0pt}{\Tdim}}}
\def\jspan{{\setbox0=\hbox{j}%
\Tdim\ht0\advance\Tdim\dp0\rule[-\dp0]{0pt}{\Tdim}}}
\def\Tspan#1{{\setbox0=\hbox{#1}%
\Tdim\ht0\advance\Tdim\dp0\advance\Tdim.55ex\rule[-\dp0]{0pt}{\Tdim}\box0}}

\def\be{\begin{eqnarray}}
\def\ben{\begin{eqnarray*}}
\def\ee{\end{eqnarray}}
\def\een{\end{eqnarray*}}

\def\=:{=\hspace{-.7em}\raisebox{1.1ex}{.}\hspace{.1em}\raisebox{-0.2ex}{.} }

\newcommand {\beq}{\begin{eqnarray}}
\newcommand {\eeq}{\end{eqnarray}}

\begin{document}

\preprint{RIKEN-MP-15}
\preprint{CCTP-2011-08}

\title{
Nucleus from String Theory}

\author{Koji {\sc Hashimoto}}\email[]{koji(at)riken.jp}
\affiliation{
{\it Mathematical Physics Lab., RIKEN Nishina Center, Saitama 351-0198,
Japan }}

\author{Takeshi {\sc Morita}}\email[]{takeshi(at)physics.uoc.gr}
\affiliation{
{\it Crete Center for Theoretical Physics, Department of Physics, University of Crete, 
 71003 Heraklion, Greece}}

\begin{abstract}
In generic holographic QCD, we find that baryons are bound to form a nucleus, and that
its radius obeys the empirically-known mass number ($A$) dependence $r\propto A^{1/3}$
for large $A$.
Our result is robust, 
since we use only a generic property of
D-brane actions in string theory. 
We also
show that
nucleons are bound completely in a finite volume.
Furthermore, 
employing 
a concrete holographic model (derived by Hashimoto, Iizuka, and Yi, 
describing a multi-baryon system in the Sakai-Sugimoto model), 
the nuclear radius is evaluated as 
${\cal O}(1)\times A^{1/3}$ [fm], which is consistent with experiments.
\end{abstract}

\maketitle


To describe atomic nuclei directly by strongly coupled quark dynamics, QCD, 
is a long-standing problem in nuclear physics and particle physics. 
It was only recent that lattice QCD simulations reproduce qualitatively the nuclear forces.
Recent progress in solving strongly coupled gauge theories 
with a new mathematical tool of superstring theory, the AdS/CFT correspondence, 
has been proven to be truly powerful in application to QCD (called holographic QCD). 

In this letter, we show by quite a generic argument in superstring theory and the AdS/CFT correspondence that 
non-supersymmetric QCD-like theories in the large $N_c$ limit host nuclei, multi-baryon bound states.
Furthermore, we can show also that the resultant nuclei have the important nuclear property 
in the real world: 
Finiteness of the nuclear size, and its mass-number dependence.
That is, 
the radius of the holographically realized nuclei is shown to be proportional to $A^{1/3}$ where $A$ is the 
mass number (the baryon number) of the nucleus.

In deriving these, we do not rely on any specific model of holographic QCD. 
What we use is only
the following two known facts: (i) Baryons are D-branes in any gravity dual of QCD-like gauge 
theories \cite{Witten:1998xy,Gross:1998gk}, 
(ii) D-brane effective actions are a dimensionally reduced Yang-Mills (YM) theory \cite{Witten:1995im}.
From these two, the formation of the nuclei and the 
mass-number dependence of the nuclear size 
follow. 
Therefore our finding is quite robust and universal for 
any holographic description of non-supersymmetric 
QCD-like gauge theories, at the large $N_c$ and at the strong coupling.

Our derivation is divided into two steps. 
\vspace*{-2mm}
\begin{itemize}
\item[1.] The system with a large number of baryons in generic holographic QCD is described effectively
by a simple bosonic matrix quantum mechanics. 
It is a pure YM action dimensionally reduced to 1 dimension \footnote{This 
was first described in \cite{Hashimoto:2008jq}, and
a precise matrix action in the realistic holographic QCD model was derived in 
\cite{Hashimoto:2010je}.},
\begin{eqnarray}
S =  c \int \! dt \; {\rm tr}_A  \left[
\frac12 (D_t X^I)^2 + \frac{g^2}{4}[X^I, X^J]^2
\right] \, .
\label{matrixaction}
\end{eqnarray}
where $I=1,\cdots,D$. 
The eigenvalues of the $A\times A$ matrix $X^i$ ($i=1,2,3$) are location of the $A$ baryons in our space,
and $X^{\hat{i}}$ $(\hat{i}=4,\cdots,D)$ is for holographic directions.
\vspace*{-2mm}
\item[2.] The system allows a non-perturbative vacuum at which the eigenvalues of 
$X^i$ form a ball-like distribution, which is nothing but a nucleus. 
The size shows the mass-number dependence $A^{1/3}$.
\end{itemize}
In the following, we shall show 1 and 2 in turn. 
Finally we present the explicit form of the nuclear density 
distribution (\ref{dist}). Together with the explicit matrix model
\cite{Hashimoto:2010je} where input 
parameters are only the $\rho$ meson mass and the NN$\pi$ coupling, we obtain the nuclear radius 
$R \sim {\cal O} (1) \times A^{1/3}$ [fm], which is consistent with the standard experimental observation.

\vspace{2mm}
\noindent
{\bf Multi-baryons described by matrices.} --- First, let us show that the system of $A$ baryons in generic 
holographic QCD is dictated by the matrix quantum mechanics (\ref{matrixaction}). 
This can be provided by known facts (i) and (ii) which we explain below.

\vspace{1mm}

(i) {\it Baryons are D-branes in any gravity dual of QCD-like gauge theories}, basically 
because baryons in large $N_c$ 
gauge theories are heavy as their mass diverges as ${\cal O}(N_c)$ while D-branes 
are solitons of string theory and thus heavy. In fact, from the point of view of baryon 
charges, it is known that, in the gravity dual side of strongly coupled gauge theories, 
D-branes wrapping compact cycles can be identified as baryons \cite{Witten:1998xy,Gross:1998gk}.
This is natural since in large $N_c$ QCD baryons appear as 
solitons of meson 
effective field theory, as in the famous Skyrme model.

\begin{figure}[t]
\begin{center}
\includegraphics[scale = 0.25, bb=200 200 400 650]{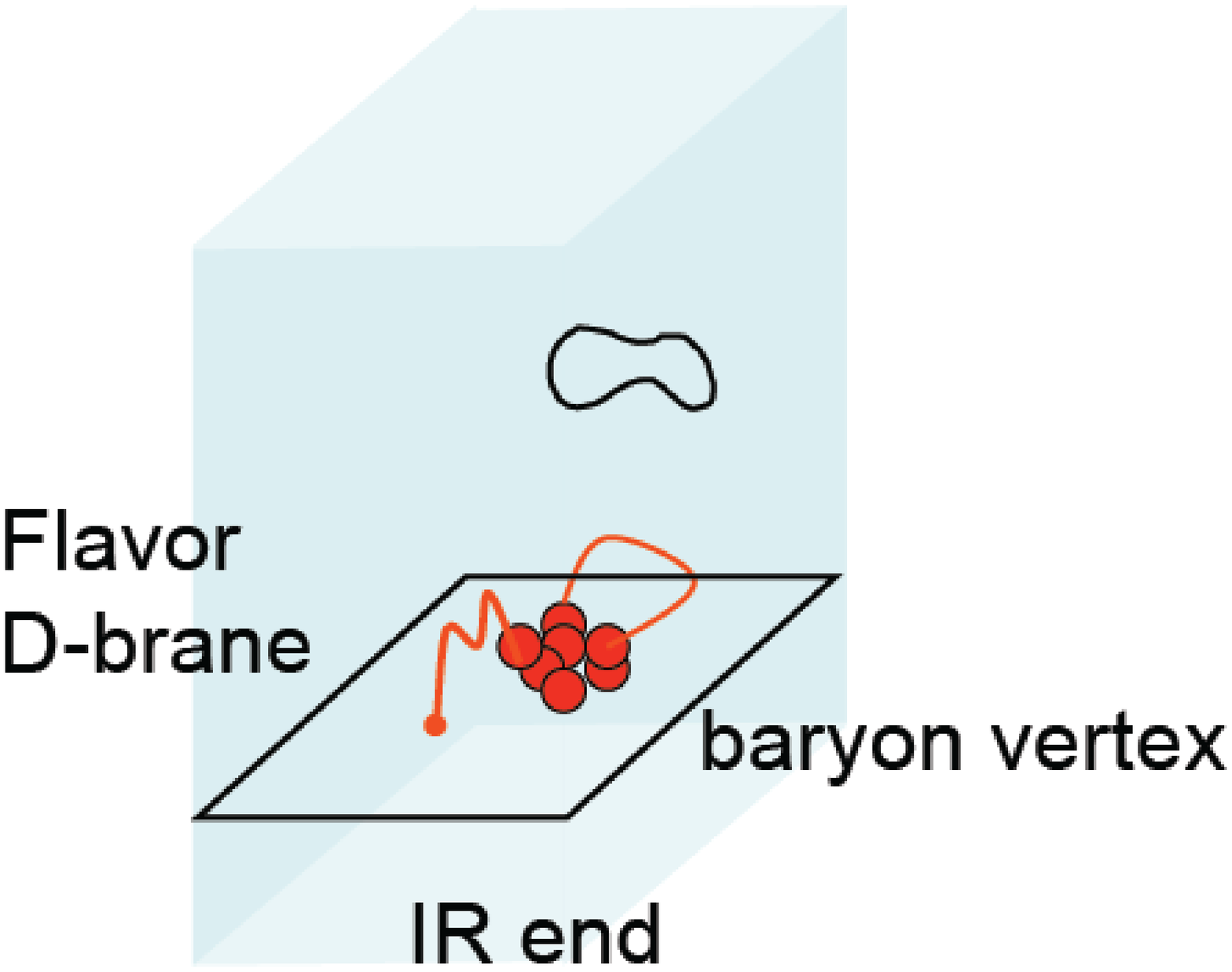}
\caption{{\footnotesize 
Gravity dual of QCD-like confining gauge theories with $A$ baryons. 
The horizontal directions are our 3-dimensional space, and the vertical direction 
is the holographic direction. This 10-dimensional spacetime is curved, 
and the bottom is capped by a smooth compact manifold, which is the IR end. 
The baryon vertices ($A$ red balls) sit on the $N_f$ flavor D-branes. 
The degrees of freedom of the baryons are open strings attached to 
the baryon vertices.}}
\label{figgeometry}
\end{center}
\vspace{-8mm}
\end{figure}

To show the universality of the statement, let us illustrate this fact with a specific and concrete 
dual gravity background provided by Witten \cite{Witten:1998zw}.
The spacetime ends smoothly at an IR end capped with non-contractable $S^4$. 
A D-brane corresponding to
a baryon is the one wrapping it (note that it is orthogonal to our spacetime
so the D-brane looks as a point particle in our space). 
As the $S^4$ is supported by $N_c$
units of a Ramond-Ramond (RR) flux in the supergravity solution of the geometry, the wrapped 
D-brane, called a baryon vertex, has to attach the ends of  $N_c$ fundamental strings,
thus attain a single baryon charge (= $N_c$ quark charge) of the gauge 
theory \cite{Witten:1998xy}. 
This situation is universal in holographic QCD, by the following two reasons.
First,  it is expected that, in any gravity dual of confining gauge theories, 
the geometry possesses a 
smooth IR end accompanying a compact manifold, known as the name of 
``confining geometry"\footnote{
Most of known examples are in this category. The D3-D(-1) background \cite{Liu:1999fc} 
is an exception where distributed D(-1)-branes
play the role of the IR end, resulting in a similar effect on baryon vertices.}. 
This end is necessary to provide a finite QCD string tension. Second,
the compact manifold always carries the $N_c$ RR flux
because 
the geometry is a gravity dual of $SU(N_c)$ gauge theories realized by $N_c$ D-branes.

Any nucleus is labeled by spins and isospins, so we need to assign flavor (isospin) charges on 
the baryon in the AdS/CFT. Here we 
review the standard holographic description of the flavor symmetry. 
The quark sector of QCD-like gauge theories is realized in the AdS/CFT correspondence by introducing 
flavor D-branes \cite{Karch:2002sh}. 
Typically we need $N_f$ D-branes which extend also along our spacetime directions,
where $N_f$ is the number of quark flavors.
The $N_c$ fundamental string emanating from the baryon vertex end on these flavor branes. The excitation of
these strings represent the flavor charges of the baryons.
This is nicely realized in Sakai-Sugimoto model of holographic QCD 
\cite{Sakai:2004cn}. Quantum excitations of the strings
give rise to the spins and isospins \cite{Hashimoto:2010je} \footnote{In the Sakai-Sugimoto model, 
the baryon vertex is absorbed into the flavor D-brane \cite{Seki:2008mu} and can be thought of as a soliton solution
on the flavor D-brane effective field theory \cite{Hata:2007mb,Hong:2007kx}. The soliton is a generalization of the
Skyrmion, and static properties of baryons can be analyzed \cite{Hashimoto:2008zw,Hashimoto:2009ys} 
(see also \cite{Hong:2007ay,Hong:2007dq}).}. In Fig.~\ref{figgeometry}, we show a typical brane configuration of
gravity dual of confining gauge theories with $A$ baryons.

\vspace{1mm}

(ii) {\it Effective theory on the 
$A$ baryon vertices for large $A$ is given by the matrix quantum mechanics}
(\ref{matrixaction}). Fundamental  
degrees of freedom on any collection of D-branes are strings connecting
the D-branes. In our case, we have $A$ baryon vertices and $N_f$ flavor D-branes, thus
we have $A\times A$ matrix $X^I$ from strings among the $A$ D-branes, in addition
to $A\times N_f$ matrix $w$ coming from strings connecting the baryon vertex and the flavor D-brane. 
These strings are shown in Fig.~\ref{figgeometry}. The effective 
theory is a $U(A)$ gauge theory, where $X^I$ is 
in the adjoint representation while $w$ is in the fundamental representation. The index $I$ runs for our spatial
directions $i=1,2,3$, and the holographic directions $\hat{i}=4,\cdots,D$. 

Our interest is  a large $A$, that is, a heavy nucleus, as $A$ is the mass number.
At the leading order in the large $A$ expansion, the fundamental representation field $w$ drops off. 
It is known in string theory that 
the $X^I$ part has a universal effective action which is a dimensionally reduced
YM action, (\ref{matrixaction}).

We ignore the gauge field components along the compact manifold
($S^4$ in the previous example), since those directions are irrelevant to
QCD \footnote{This is standard for any brane model of holographic QCD, though no
consistent truncation has been known.}. 
The time component gauge field $A_t$ still remains, 
but it is not dynamical and just ensures the local gauge invariance \footnote{
A 1-dimensional Chern-Simons term $\int dt \; {\rm tr} A_t$ in the action contributes only
the overall $U(1)$ sector of the $U(A)$, thus is a sub-leading order in the large $A$ expansion.}.
Fermionic superpartners are heavy due to the supersymmetry breaking and assumed not effective.

An explicit action can be derived once we fix the  species of the branes in holographic QCD.
The nuclear matrix model of \cite{Hashimoto:2010je} \footnote{The field $X^5$ was not written in 
\cite{Hashimoto:2010je}, but together with $X^4$ they span the transverse cigar geometry. } 
was derived in the Witten's geometry
with the flavor D8-branes of the Sakai-Sugimoto model, 
\begin{align}
S &=\frac{\lambda N_c M_{\rm KK}}{3^3 \pi}
\int \! dt \; {\rm tr}_A 
\left(
\frac12 (D_t X^I)^2
+ \frac{\lambda^2 M_{\rm KK}^4}{3^6 \pi^2}
[X^I,X^J]^2
\right.
\nonumber \\
&\left.
-\frac{1}{3}M_{\rm KK}^2 ((X^4)^2+(X^5)^2)
\right)
+ \mbox{sub-leading in $1/A$}.
\label{nmm}
\end{align}
Here $I$ runs from 1 to $D=5$, $\lambda \equiv N_c g_{\rm QCD}^2$ is the QCD 'tHooft coupling,
and $M_{\rm KK}$ is a dynamical scale (roughly corresponding to the QCD scale).
With a rescaling $Y^I \equiv X^I (\lambda N_c M_{\rm KK}/3^3\pi)^{1/2}$,
the leading order is written in a canonical expression,
\begin{eqnarray}
S = \!\int \!\! dt
\; {\rm tr}_A \!
\left(
\frac12 (D_t Y^I)^2 - \frac{m_Y^2}{2} (Y^{\hat{i}})^2
+\frac{g_0^2}{4}[Y^I, Y^J]^2\!
\right), \; 
\label{normaction}
\end{eqnarray}
where the matrix model coupling squared is $g_0^2 \equiv 2^2\lambda M_{\rm KK}^3/(3^3 \pi N_c)$.
 In comparison to (\ref{matrixaction}) here 
 there is a mass term with $m_Y^2 = (2/3) M_{\rm KK}^2$, but
it is qualitatively irrelevant in our analysis, see Footnote [32]. 
The mass term is only along the holographic direction $X^{\hat{i}}$,
which means that the baryon vertex is stable at the IR end of the geometry $X^{\hat{i}}=0$.


\vspace{2mm}
\noindent
{\bf Formation of nucleus and 
nuclear size.}
 --- We shall show that indeed the eigenvalues of 
the matrix $X^I$ are bound to each other, which directly means a formation of a nucleus, since the eigenvalues
are the location of the $A$ baryons.

As the action (\ref{matrixaction}) is a dimensional reduction of a YM theory,
we can apply the generic argument by L\"uscher \cite{Luscher:1982ma} on spectra of YM theory on torii.
L\"uscher proved a theorem that the ground state of the theory (\ref{matrixaction}) has eigenvalues
bound to a finite region, and is
invariant under the $U(A)$ rotation and the spatial rotation. Even though there exists a flat 
direction of the potential at which all $X^I$ are diagonal, the eigenvalues are bound
because the flat direction is narrow for large values of $X^I$ and the 
quantum dynamics suppresses percolation. So, {\it this theorem ensures a formation
of a spherical nucleus in generic holographic QCD}.

Next, we show 
the nuclear radius $\propto A^{1/3}$. It follows
simply from a dimensional analysis of the action (\ref{matrixaction}).
The nuclear size is given by the distribution of the entries of $X^i$ ($i=1,2,3$), so we can define the mean square radius as 
\begin{eqnarray}
r_{\rm mean}^2 \equiv \frac{1}{A} \sum_{i=1}^3
\left\langle
{\rm tr}_A [X^i X^i]
\right\rangle.
\label{r-mean}
\end{eqnarray}
The expectation value is taken with the ground state of the matrix quantum mechanics.
The normalization $1/A$ is understood from the case of diagonal $X$ with which
individual location of the baryons makes sense.
Using the properly normalized action as in (\ref{normaction}), we can estimate
\begin{eqnarray}
 \frac{1}{A^2} 
\sum_{i=1}^3 \left\langle
{\rm tr}_A [Y^i Y^i]
\right\rangle=c_0 \lambda_A^{-1/3} + O(1/A)  ,
\label{sat}
\end{eqnarray}
at large $A$,
where $\lambda_A \equiv A g_0^2$ is a 'tHooft coupling of the quantum mechanics and
 $c_0$ is an undetermined $A$-independent dimensionless constant.
We have used a standard 'tHooft expansion \cite{'tHooft:1973jz}. 
The ${\lambda}_A $ dependence was determined from the mass dimension: 
$[\lambda_A]=3$ and $[Y^i]=-1/2$.
($\lambda_A$ is the unique dimension-ful parameter in the action.)
Since the rescaling from $X$ to $Y$ is $A$-independent,
we obtain $\sqrt{r^2_{mean}} \propto  A^{1/3}$.
{\it Therefore we conclude that in generic holographic
QCD a nucleus forms as a bound state of $A$ baryons at large $A$, 
and exhibits the 
correct $A$ dependence of the nuclear size.}


\vspace*{2mm}

\noindent
{\bf Nuclear density distribution.} --- A certain approximation of the matrix quantum mechanics 
(\ref{matrixaction}) enables us to compute more detailed information of the bound state: nuclear density 
distribution. 
Here we shall show that the nuclear density computed by the matrix quantum mechanics 
(\ref{matrixaction}) 
vanishes outside a certain radius, which is the finiteness of nuclei.

To this end, we employ a RR density formula \cite{Taylor:1999gq} in a $D$ dimensional spacetime,
at the leading order in $1/A$, developed in the context of 
Matrix theory \cite{Banks:1996vh} in superstring theory:
\begin{eqnarray}
\rho(x) = \frac{1}{(2\pi)^D}
\int \! d^Dk \; e^{-i k\cdot x} \left\langle {\rm tr}_A
\exp [i k\cdot X]\right\rangle. 
\label{exp}
\end{eqnarray}
Since our baryon vertices are RR-charged D-branes, 
(\ref{exp}) is equivalent to the baryon
charge distribution. We evaluate the expectation value at zero temperature \footnote{
The contribution of the $Y^I$ is dominant at the zero temperature.
However, since the model (\ref{normaction}) would be in a confinement phase in low temperature \cite{Mandal:2009vz},
the thermal excitation of $Y^I$ is suppressed. 
Thus the $1/A$ corrections of the model would be relevant in a finite temperature. }.

To evaluate (\ref{exp}) with the model (\ref{matrixaction}), we need a help of a large $D$
expansion of matrix models \cite{Hotta:1998en,Mandal:2009vz,Mandal:2011hb}, where 
$D$ counts the number of the matrices $X^I$ ($I=1,\cdots, D$). We fix 
$\tilde{\lambda}_A \equiv g_0^2 A D$ to be finite,
and take the large $D$ and large $A$ limits. The large $D$ limit is known to be a good approximation
as discussed in \cite{Mandal:2009vz} even for small $D$ $(\ge 2)$ qualitatively. 
According to \cite{Mandal:2009vz}, at the leading order of the expansion,
we obtain a non-perturbative vacuum at zero temperature, which is characterized by
\begin{eqnarray}
\frac{1}{A^2} \sum_{I=1}^D \left\langle {\rm tr}_A [Y^I Y^I] \right\rangle_{T=0}  = \frac{D}{2} \tilde{\lambda}^{-1/3}_A.
\label{character}
\end{eqnarray}
This indeed is consistent with (\ref{sat}).
Around this vacuum, $Y^I$ behaves as a free massive field 
with a mass $\tilde{\lambda}^{1/3}_A $, as the interaction is suppressed by $1/D$ \footnote{\label{last-foot}
If the dynamical mass $\tilde{\lambda}_A^{1/3}$ is larger than $m_Y$, 
we can ignore the effects of $m_Y$ in (\ref{normaction}). This is the case in our model (\ref{nmm}).
On the other hand, if $m_Y$ is large enough,
one can integrate out $X^{\hat{i}}$ and again the effects can be ignored by effectively treating $D=3$.}.
By using the propagator of the free massive scalar at zero temperature,
\begin{eqnarray}
\langle Y^I_{ab}(t) Y^J_{cd}(0)\rangle_{T=0} \sim \frac{1}{2\tilde{\lambda}^{1/3}_A} 
\delta_{ad} \delta_{bc}\delta^{IJ}
\quad (t \sim 0) ,
\label{2pt}
\end{eqnarray}
 we can evaluate (\ref{exp}) as 
\begin{eqnarray}
\langle {\rm tr}_A \exp (ik\cdot Y)\rangle_{T=0}
&=&\sum_{n=0}^\infty \frac{A}{n!(n+1)!}\left(\frac{-Ak^2}{2 \tilde{\lambda}^{1/3}_A}    \right)^{n} \nonumber \\
  &=& A \frac{2}{r_0 |k|}J_1(r_0 |k|)
\end{eqnarray}
where $r_0 \equiv (2A/\tilde{\lambda}_A^{1/3})^{1/2}$. 
Here we have used a fact that only ladder diagrams \cite{Erickson:2000af} contribute to (\ref{exp}).
The distribution in our 3-dimensional space can be obtained by a Fourier transform followed by  
an integration over the holographic directions,
\begin{align}
\rho(r)&= 
\int \prod_{\hat{i}=4}^{D}dx^{\hat{i}} \frac{1}{(2\pi)^D}
\!\!\int \! d^Dk \; e^{-ik\cdot x} \langle {\rm tr}_A \exp (ik\cdot Y)\rangle_{T=0}
\nonumber \\
 &=\left\{
\begin{array}{ll}
\displaystyle\frac{A}{\pi^2 r_0^2 \sqrt{r_0^2-r^2}} & (r<r_0)
\\[10pt]
0 & (r_0 < r)
\end{array}
\right.
\label{dist}
\end{align}
where $r$ is a radial coordinate in the 3-dimensional space spanned by 
$\{Y^1, Y^2, Y^3\}$. 
{\it We found that nuclear density is zero outside a cetrain radius, which is the finiteness
of nuclei.}

\begin{figure}[t]
\begin{center}
\begin{minipage}{5cm}
\includegraphics[scale=0.2, bb=160 220 400 400]{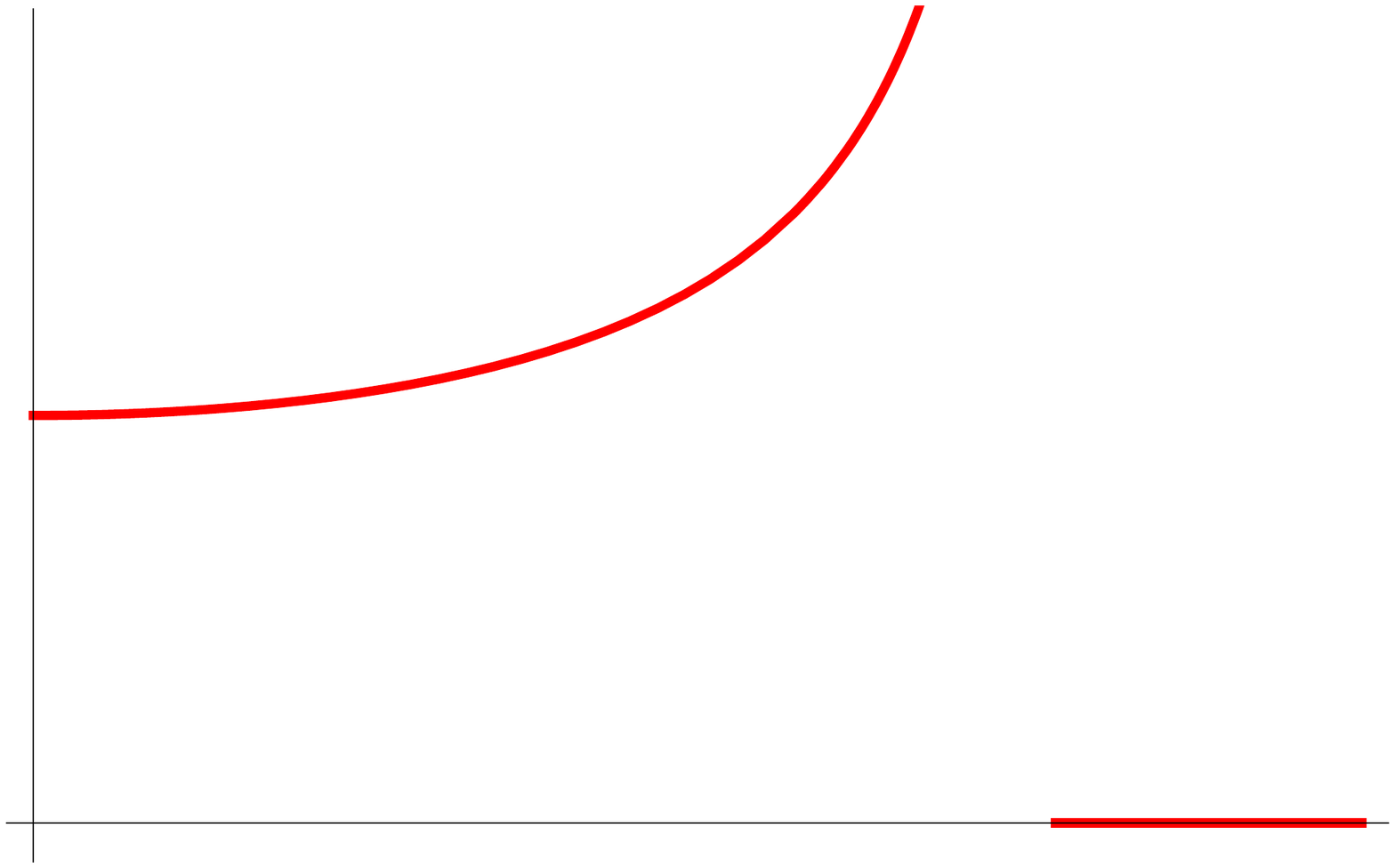}
\put(40,-5){\small{$r$}}
\put(-100,75){\small{$\rho(r)$}}
\put(0,-5){\small{$r_0$}}
\put(-100,40){\small{$\frac{A}{\pi^2 r_0^3}$}}
\caption{{\footnotesize The nuclear density distribution $\rho(r)$ 
obtained in the large $D$ limit, (\ref{dist}). At the core of the nucleus 
$r\sim 0$, the distribution is homogeneous. For $r>r_0$, the density is 
exactly zero, meaning the complete finiteness of nuclei. A graphical image
is drawn in the next figure.}}
\label{dens}
\end{minipage}
\hspace{2mm}
\begin{minipage}{3cm}
\includegraphics[scale=0.12, bb=250 100 300 600]{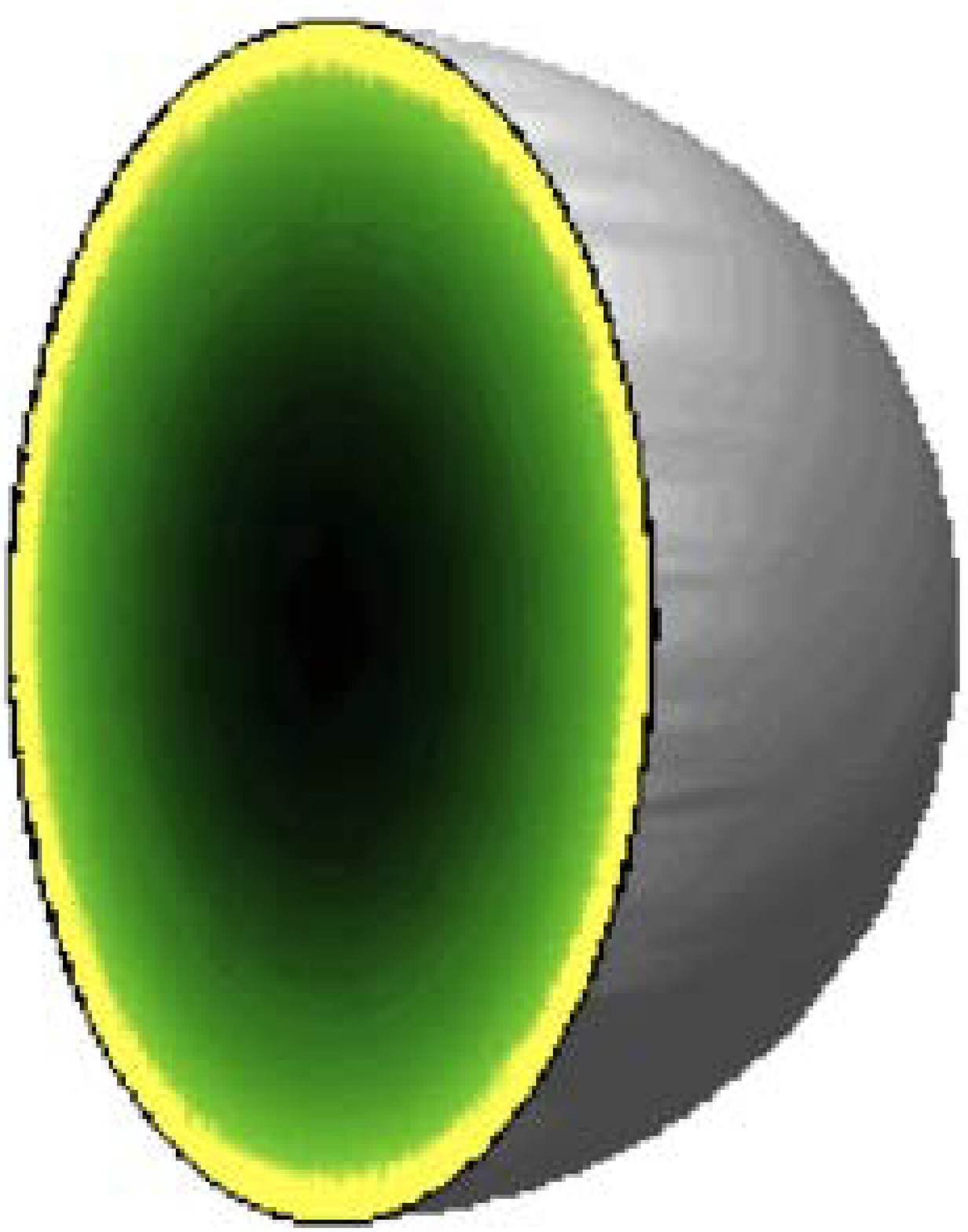}
\caption{{\footnotesize The density distribution of
the nucleus. Darker color means smaller density. Ignoring the
surface defect, we find a homogeneous 
density distribution.}}
\label{dens2}
\end{minipage}
\end{center}
\vspace{-8mm}
\end{figure}

Plotting this function shows the density distribution
given in Fig.~\ref{dens} and Fig.~\ref{dens2}. 
We notice that at $r=r_0$ the density goes up, 
which is not really the case for realistic nuclei. 
However, since this surface part is at a sub-leading in the large $A$ 
expansion, explicit computations of $1/A$ corrections may be necessary to see the details.
Other possible origin may be the formula \eqref{exp} itself \cite{Hashimoto:2004fa}. 
We hope to come back to this issue in
a near future.

We also notice that the isospin structure of the nucleus is invisible, since it is encoded in $w$, which is sub-dominant in the large $A$ expansion. In contrast, the distribution in $X^{\hat{i}}$ directions indicates that the nucleus includes components of excited baryon resonances.


\vspace*{2mm}

\noindent
{\bf Evaluation of nuclear size.} --- Finally we shall study numerics. Using the explicit matrix quantum mechanics (\ref{nmm}) of \cite{Hashimoto:2010je}, we can calculate (\ref{r-mean}) through (\ref{2pt}) as
\begin{eqnarray}
\left. \sqrt{r^2_{\rm mean}} \right|_{T=0}= \frac{3^{5/2} \pi^{2/3}}{2^{5/6} 5^{1/6}}
\frac{A^{1/3}}{M_{\rm KK} (N_c \lambda^2)^{1/3}}.
\label{radsmall}
\end{eqnarray}
A natural choice of the experimental inputs 
should be $g_{\pi NN}\sim13.2$ and $m_\rho\sim 776$ [MeV] as these dictate nuclear forces (the model 
\cite{Hashimoto:2010je} is in the chiral limit).
From the work \cite{Hashimoto:2008zw} we find that these experimental inputs
are reproduced when $\lambda=5.31$ and $M_{\rm KK}=949$ [MeV]. This input leads us to
a nuclear mean square radius
\begin{eqnarray}
\sqrt{r^2_{\rm mean}}\sim 0.7 A^{1/3} \; {\rm [fm]}.
\end{eqnarray}
As an order estimate, this is quite close to the observed value 
$\sqrt{r^2_{\rm mean}} \simeq 1.0 A^{1/3}$ [fm] (which is equivalent to the well-known nuclear radius 
$R \simeq 1.2 A^{1/3}$ [fm] as 
$r^2_{\rm mean}= (3/5) R^2$ for a homogeneous nuclear density.)

One can alternatively think of $r_0$ in 
the density distribution 
(\ref{dist}) as the nuclear radius.
After going back from $Y$ to $X$, with the same experimental values as the inputs, we obtain
$r_0 \sim 0.8 \times A^{1/3}$ [fm]. This again is consistent with experiments.

\noindent
{\bf Acknowledgment}: 
K.H.~is grateful to Norihiro Iizuka for valuable discussions, and would like to thank 
H.~Toki, A.~Hosaka and T.~Hatsuda for insightful comments. 
T.M.~would like to thank Gautam Mandal for valuable discussions.
T.M.~would also like to thank KEK for
hospitality where part of the work was done.
K.H.~is partly supported by
the Japan Ministry of Education, Culture, Sports, Science and
Technology.
T.M.~is partially supported by Regional Potential program of the E.U. FP7-REGPOT-2008-1: CreteHEPCosmo-228644 and by Marie Curie contract
PIRG06-GA-2009-256487.




\begin{thebibliography}{99}


\bibitem{Hashimoto:2010je}
  K.~Hashimoto, N.~Iizuka, P.~Yi,
  JHEP {\bf 1010}, 003 (2010).
  [arXiv:1003.4988 [hep-th]].

\bibitem{Witten:1998xy}
  E.~Witten,
  JHEP {\bf 9807}, 006 (1998).
  [hep-th/9805112].  

\bibitem{Gross:1998gk}
  D.~J.~Gross, H.~Ooguri,
  Phys.\ Rev.\  {\bf D58}, 106002 (1998).
  [hep-th/9805129].
  
\bibitem{Witten:1995im}
  E.~Witten,
  Nucl.\ Phys.\  B {\bf 460}, 335 (1996)
  [arXiv:hep-th/9510135].

\bibitem{Hashimoto:2008jq}
  K.~Hashimoto,
  Prog.\ Theor.\ Phys.\  {\bf 121}, 241 (2009)
  [arXiv:0809.3141 [hep-th]];
  JHEP {\bf 0912}, 065 (2009)
  [arXiv:0910.2303 [hep-th]].
  
\bibitem{Witten:1998zw}
  E.~Witten,
  Adv.\ Theor.\ Math.\ Phys.\  {\bf 2}, 505-532 (1998).
  [hep-th/9803131].
  
\bibitem{Liu:1999fc}
  H.~Liu, A.~A.~Tseytlin,
  Nucl.\ Phys.\  {\bf B553}, 231-249 (1999).
  [hep-th/9903091].  
  
\bibitem{Karch:2002sh}
  A.~Karch, E.~Katz,
  JHEP {\bf 0206}, 043 (2002).
  [hep-th/0205236].  
  
\bibitem{Sakai:2004cn}
  T.~Sakai, S.~Sugimoto,
  Prog.\ Theor.\ Phys.\  {\bf 113}, 843-882 (2005).
  [hep-th/0412141].
  
\bibitem{Seki:2008mu}
  S.~Seki, J.~Sonnenschein,
  JHEP {\bf 0901}, 053 (2009).
  [arXiv:0810.1633 [hep-th]].  
  
\bibitem{Hata:2007mb}
  H.~Hata, T.~Sakai, S.~Sugimoto, S.~Yamato,
  Prog.\ Theor.\ Phys.\  {\bf 117}, 1157 (2007).
  [hep-th/0701280 [HEP-TH]].

\bibitem{Hong:2007kx}
  D.~K.~Hong, M.~Rho, H.~-U.~Yee, P.~Yi,
  Phys.\ Rev.\  {\bf D76}, 061901 (2007).
  [hep-th/0701276 [HEP-TH]].
  
\bibitem{Hashimoto:2008zw}
  K.~Hashimoto, T.~Sakai, S.~Sugimoto,
  Prog.\ Theor.\ Phys.\  {\bf 120}, 1093-1137 (2008).
  [arXiv:0806.3122 [hep-th]].  
  
\bibitem{Hashimoto:2009ys}
  K.~Hashimoto, T.~Sakai, S.~Sugimoto,
  Prog.\ Theor.\ Phys.\  {\bf 122}, 427-476 (2009).
  [arXiv:0901.4449 [hep-th]].  
  
\bibitem{Hong:2007ay}
  D.~K.~Hong, M.~Rho, H.~-U.~Yee, P.~Yi,
  JHEP {\bf 0709}, 063 (2007).
  [arXiv:0705.2632 [hep-th]].  
  
\bibitem{Hong:2007dq}
  D.~K.~Hong, M.~Rho, H.~-U.~Yee, P.~Yi,
  Phys.\ Rev.\  {\bf D77}, 014030 (2008).
  [arXiv:0710.4615 [hep-ph]].  
  
\bibitem{Luscher:1982ma}
  M.~Luscher,
  Nucl.\ Phys.\  {\bf B219}, 233-261 (1983).  
  
\bibitem{'tHooft:1973jz}
  G.~'t Hooft,
  Nucl.\ Phys.\  B {\bf 72}, 461 (1974).

  
\bibitem{Taylor:1999gq}
  W.~Taylor, M.~Van Raamsdonk,
  Nucl.\ Phys.\  {\bf B558}, 63-95 (1999).
  [hep-th/9904095].
  
\bibitem{Banks:1996vh}
  T.~Banks, W.~Fischler, S.~H.~Shenker, L.~Susskind,
  Phys.\ Rev.\  {\bf D55}, 5112-5128 (1997).
  [hep-th/9610043].  

\bibitem{Hotta:1998en}
  T.~Hotta, J.~Nishimura, A.~Tsuchiya,
  Nucl.\ Phys.\  {\bf B545}, 543-575 (1999).
  [hep-th/9811220].
  
\bibitem{Mandal:2009vz}
  G.~Mandal, M.~Mahato, T.~Morita,
  JHEP {\bf 1002}, 034 (2010).
  [arXiv:0910.4526 [hep-th]].  
  
\bibitem{Mandal:2011hb}
  G.~Mandal and T.~Morita,
  arXiv:1103.1558 [hep-th].

  
\bibitem{Erickson:2000af}
  J.~K.~Erickson, G.~W.~Semenoff and K.~Zarembo,
  Nucl.\ Phys.\  B {\bf 582}, 155 (2000)
  [arXiv:hep-th/0003055].
  
\bibitem{Hashimoto:2004fa}
  K.~Hashimoto,
  JHEP {\bf 0404}, 004(2004).[hep-th/0401043].
  \end{thebibliography}
\end{document}